\documentclass[aps,12pt]{revtex4}

\usepackage{graphicx}
\usepackage{indentfirst}

\begin{document}

\title{Heat transfer between a nano-tip and a surface}

\author{Pierre-Olivier Chapuis$^{1}$\footnote{Electronic mail : olivier.chapuis@em2c.ecp.fr}, Jean-Jacques Greffet$^{1}$, Karl Joulain$^{2}$ and Sebastian Volz$^{1}$\footnote{E-mail : sebastian.volz@em2c.ecp.fr}}
\affiliation{$^{1}$Laboratoire d'Energ\'etique Mol\'eculaire et Macroscopique, Combustion\\CNRS UPR 288, Ecole Centrale Paris\\Grande Voie des Vignes, F-92295 Ch\^atenay-Malabry cedex, France\\
$^{2}$Laboratoire d'Etudes Thermiques\\
CNRS UMR 6608 and ENSMA,
BP 40109, Futuroscope, \\
F-86961 Chasseneuil cedex, France  
}

\linespread{1}

\begin{abstract}

We study quasi-ballistic heat transfer through air between a hot nanometer-scale tip and a sample. The hot tip/surface configuration is widely used to perform non-intrusive confined heating. Using a Monte-Carlo simulation, we find that the thermal conductance reaches 0.8~MW.m$^{-2}$K$^{-1}$ on the surface under the tip and show the shape of the heat flux density distribution (nanometer-scale thermal spot). These results show that a surface can be efficiently heated locally without contact. The temporal resolution of the heat transfer is a few tens of picoseconds.
\end{abstract}
\maketitle
\
\
\
\
\
\
\
\
\
\
\
Understanding the heat transfer between a hot tip and a substrate is a key issue in nanometer-scale devices operating at high temperatures such as scanning thermal microscopy (SThM)~\cite{Wickramasinghe}\cite{Dinwiddie} and also thermally assisted data storage~\cite{Millipede}. Indeed, several of the future methods to write or read information are based on the use of heat transfer between an Atomic Force Microscope (AFM) probe and the disk, because a spatial resolution below $100$~nm can be achieved due to the small size of the tip~\cite{Vettiger}. Contact heating can be used to write by melting the substrate~\cite{Millipede}. But the tip-sample contact is then responsible of tip damaging~\cite{Yang}. Moreover, in such devices, heat is dissipated over large distances if the hot part of the tip is micrometric~\cite{IJHMT}. A non-contact localized heating through air with a moving media could be a solution. Different methods as Joule (\emph{e.g.}~\cite{IBM}) or Peltier heating (\emph{e.g.}~\cite{Majumdar}) are possible to heat the extremity of the tip. A quantitative estimation of the heat transfer through air clearly appears as a key step to design thermally assisted data storage techniques. In this paper we report a calculation of the heat transfer between the apex of a tip and a surface. We discuss the leading mechanism, the spatial and temporal resolution.

\begin{figure}
\begin{center}
\includegraphics[width=14cm]{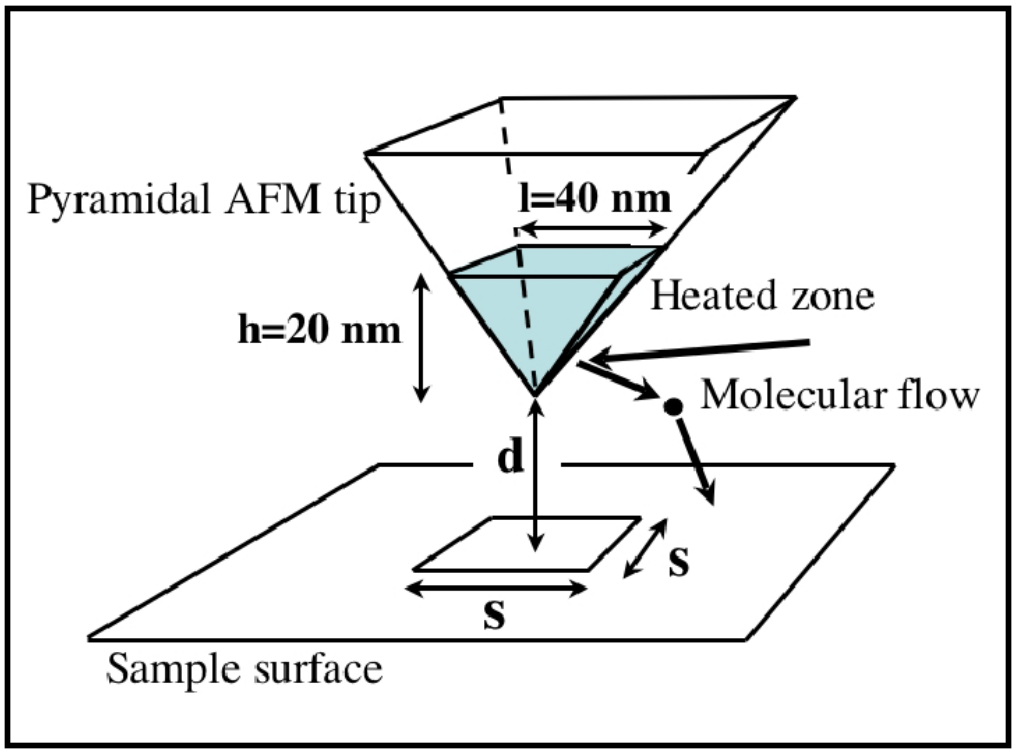}
\caption{A skip of the tip. Typical lengths used are l=40 nm and h=20 nm. The distance d is in the nanometric range.}
\end{center}
\end{figure}
 
When a hot tip and a cold sample are not in contact, heat can be exchanged through conduction \cite{Majumdar} or radiation \cite{Mulet2001}. The radiation contribution can be evaluated using the model described in ref.\cite{Mulet2001}. The heat conduction cannot be modelled by classical Fourier diffusion because the air mean free path (MFP) is on the order of the average distance between the tip and the sample. Pure ballistic transfer is neither valid : this problem has to be solved in the intermediate regime.  3D effects have also to be included to observe the impact of the tip geometry. We therefore use a Monte-Carlo approach associated with the linear response theory in order to directly solve space and time dependent problems. This model also allows to study the time dynamics of heat conduction through the air gap.

We consider the extremity of a pyramidal AFM tip with a square base of $l \times l$ with $l$=40 nm and a height $h$=20 nm. The distance $d$ between the tip and the surface is in the nanometric range. We give the thermal characteristics of heat diffusion in the tip. In the stationary regime, heat confinement remains realistic since heat diffusion in the tip generates a significant temperature decay over a few tens of nanometers. In the transient regime, the heat diffusion in the tip has a characteristic time ($h^2/a$, where a is the diffusivity of the material) on the order of $1$ ns. 

We now describe how the radiation is evaluated. The radiation between a sphere with same volume as the heated zone of the tip and the surface was calculated for different materials. For a distance $20$ nm between the center of the sphere at $800$ K and the surface of same material at $300$ K , we found a flux of $5\,10^{-17}$ W for Si and $1\,10^{-14}$ W for silver. As discussed in ref.\cite{Mulet2001}, this flux can be increased if both the tip and the surface have an infrared active optical phonon resonance. For glass, we found $9\,10^{-9}$ W. 

We now turn to the description of the conduction model. Its basis consists in calculating the time evolution of the local heat flux on the sample $<\mathrm{q}^{R}_{th} \left(r,t,T_{tip}\right)>$ due to an impulse of molecular flux leaving the tip. The heat flux response $q_{th} \left( r,t \right)$ to a given molecular flux excitation $q_{m}^{tip}\left(\tau \right)$ is then given by
\begin{equation}
q_{th} \left( r,t \right) = \int_{0}^{t} <\mathrm{q}^{R}_{th} \left(r,t-\tau ,T_{tip}\left(\tau \right)\right)>  S_{tip}.q_{m}^{tip}\left(\tau \right) d\tau
\end{equation}
where $<\mathrm{q}^{R}_{th}>$ plays the role of a susceptibility. $S_{tip}$ is the area of the heated part of the tip.

Let us now discuss how the susceptibility is computed using a Monte-Carlo approach. The heat transfer is due to air molecules which are in first approximation nitrogen molecules. A molecule is considered to carry the translational and rotational energy $\frac{5}{2} k_{B} T$  where  $k_{B}$ is the Boltzmann's constant (reference \cite{Rohsenow}). A cold molecule is heated when impacting the hot tip and then flies with a velocity chosen according to a Boltzmann's law at $T_{tip}$. The emission laws for velocity $v$ and direction described by the cylindrical angles $\theta$ and $\varphi$ are derived from the equilibrium molecular flux leaving the probe at $T_{tip}$ 
\begin{eqnarray}\nonumber
q_{m}^{tip} = & n_{tip} {\left(\frac{m}{2\pi k_{B} T_{tip} } \right) }^{3/2} \int_{v=0}^{\infty} v^{3} \exp{\left(\frac{-m v^{2}}{2 k_{B} T_{tip}}\right)} dv \\
& \times \int_{\theta=0}^{\frac{\pi}{2}} \cos \theta \sin \theta  d \theta \int_{ \varphi=0}^{2\pi}  d\varphi .
\end{eqnarray}
where $m$ is the diatomic molecular mass. $n_{tip}$ is the number of molecules per unit volume derived from the condition of null molecular flux at the tip surface. The sample and gas are at the ambiant temperature $T_{a}$. Neglecting the increase of $T_{a}$ due to the presence of the tip, the incident molecular flux is given by Eq.~(2) where $n_{tip}$ is replaced by the number of molecules per unit volume in the gas denoted $n_{g}$ and $T_{tip}$ is replaced by $T_{a}$. A molecule leaving the tip flies until undergoing a collision with another molecule or with the sample surface. The path length between two collisions is computed according to an exponential law with a characteristic decay length given by the MFP $\Lambda=\frac{1}{ \pi \sqrt{2} \left(2 R_{pot} \right)^2 n}$ which is $55$~nm at atmospheric pressure. $R_{pot}=0.2075$~nm \cite{Bird} denotes the molecular radius. The velocity of the collision partners are computed with Boltzmann's law taken at ambient temperature $T_{a}$. It allows to define a new velocity and a new direction after the collision for the test-molecule according to the Very Hard Sphere model~\cite{Bird}. A molecule may undergo several collisions before it reaches the sample surface. Hot molecules which remain in the gas are discarded and the contribution of collision partners to the tip-sample heat flux is neglected. Finally, the thermal flux received by an element $ \delta x \times \delta y$ of the surface is 
\begin{equation}
<\mathrm{q}_{th}^{R}\left(r,t,T_{tip} \right)>= \frac{E\left(r,t\right)}{\delta x \delta y \delta t N}
\end{equation}
where $N$ is the number of emitted molecules and $E\left(r,t\right)$ is the net energy transfered by the molecules emitted by the tip and falling in the area $ \delta x \delta y$ during the time interval $\delta t$. Each molecule arrives with an energy determined by the Monte-Carlo simulation and leaves the area with the mean energy $T_{a}$. We assume therefore that the kinetic energy is fully absorbed~\cite{Rohsenow}. Those molecules are considered to leave the tip at the same initial time and the arrival time $t$ is given by their flight time. The calculated susceptibility is a statistical average over $N  \sim 4 \times 10^{8}$ molecules. At $T_{tip}=800$~K, the mean power given by the tip to air molecules is about 90~nW.

\begin{figure}
\begin{center}
\includegraphics[width=9.5cm]{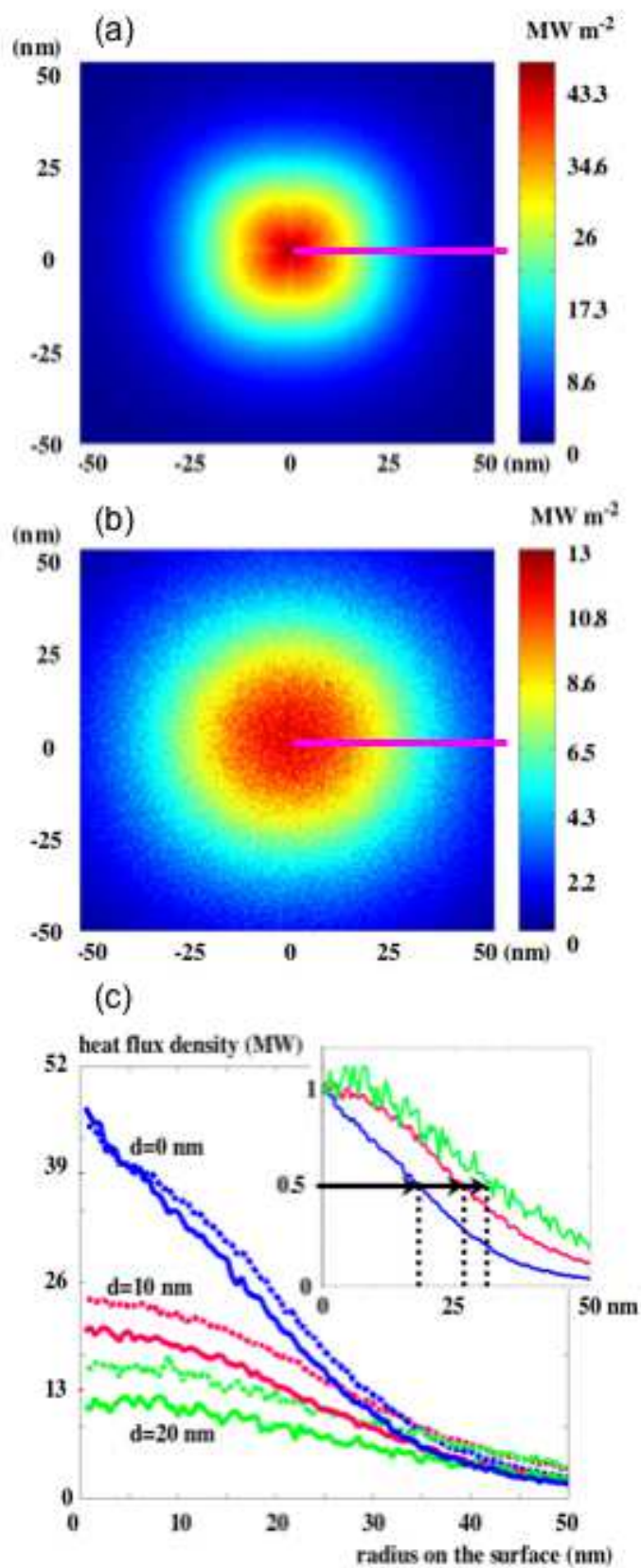}
\caption{Heat flux density on the substrate. (a) $d=0$~nm (b) $d=10$~nm (c) Heat flux density along the section shown in (a)(b). Dashed lines show results when collisions are not taken into account. The inset shows the normalized heat flux density.}

\end{center}
\end{figure}

Figures~2(a) and 2(b) represent the heat flux density deposited on the sample surface when the tip is in contact (top) and when $d=10$~nm (bottom) in a steady-state analysis. It is obtained by temporal integration of the susceptibility. The characteristic size (FWHM) of the heated zone in contact is a square of edge 35~nm. This corresponds to the best spatial resolutions for SThM and also to disk densities up to Tbit.in$^{-2}$. The geometry of the heat flux density reproduces the square shape of the tip projection in the case of contact. For heights larger than $d=10$~nm, the imprint of the tip is lost and the deposited energy becomes axisymmetric. This shows that the shape of the tip no longer influences the heat transfer. This behavior is still observed if we do not include collisions in the model. It follows that it is due to the isotropy of the velocity distribution of emitted molecules. Previous models based on vertical projection might not reproduce correctly this phenomenon under $d=10$~nm~\cite{Majumdar}. Levels of heat flux densities obtained for tip temperature of 800~K are higher than 45~MWm$^{-2}$ in contact. The values that we found correspond to a total power on the order of $10$~nW much larger than the radiation contribution when using silicon. The temperature of a $10$~nm thick film of insulator material (thermal conductivity k=1~Wm$^{-1}$K$^{-1}$) would be increased by about 80~K in 1~$\mu s$ in contact and by 25~K when $d=10$~nm. This indicates that heating through air with a moving media is feasible.
On Figure 2(c), we report the radial distribution of the heat flux density. In contact, a linear decrease is found until $35$~nm. As seen in the inset, the FWHM of the thermal spot increases by more than $10$~nm when $d$ increases from $d=0$ to $20$~nm. The comparison of dashed lines and plain lines shows that the effect of the collisions is to further reduce the heat flux. 

\begin{figure}
\begin{center}
\includegraphics[width=14cm]{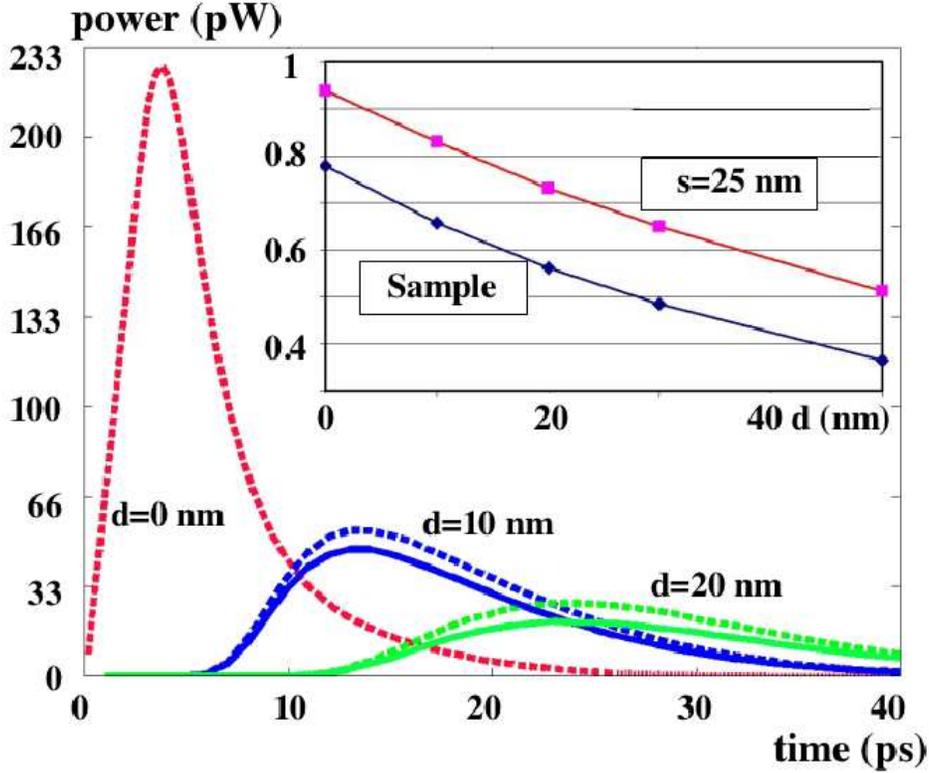}
\caption{Temporal evolution of the susceptibility : power given to a $s=15$~nm square of the substrate for a $l=40$~nm and $h=20$~nm tip. Dashed lines represent pure ballistic calculations. The inset shows the full power on the sample  normalized by the pure ballistic flux when the tip is retracted. This illustrates the decay of the flux due to collisions.} 
\end{center}
\end{figure}

Figure~3 represents the evolution of the susceptibility~\cite{MTE} $<\mathrm{q}^{R}_{th} \left(r,t, T\right)>$ integrated over a square surface of side $s=15$~nm~ under the tip with $l=40$~nm and $h=20$~nm at various heights $d$. The minimum heating time is equal to the time width of the flux response. This is on the order of $10$~ps. This is not a limiting factor for writing applications since this time is much smaller than the conduction time scale in the tip. As shown in the inset, the power given to the sample surface at $d=30$~nm is half of what is found in a pure ballistic calculation and $65\%$ for a square of edge $25$~nm. Although the susceptibilities are not qualitatively different, it is necessary to take into account effects of collisions. Note that the collisions are responsibles for the loss of $70\%$ of the power in comparison to the ballistic one when the height $d$ is equal to the mean free path.

In summary, we have studied the heat transfer through air between the hot extremity of a tip and a surface. The hot tip/surface configuration is widely used to heat locally a surface. We find that for silicon, air conduction dominates the heat transfer. The values of the heat flux density reach $0.5$~MWm$^{-2}$K$^{-1}$ at a distance of $35$~nm from the axis of the tip. The time scale of the conduction heat transfer is on the order of tens of picoseconds. These figures show that heat transfer through the air at distances on the order of $10$~nm is sufficient to allow non-contact scanning thermal microscopy or thermally assisted data storage.

\section*{Acknowledgments}

This work was supported by a grant of the French Ministry of Research (ACI Surfaces/Interfaces). We thank J.P. Nozi\`eres for helpful discussions.

{
\small
\linespread{1}

}


\begin{thebibliography}{99}

\bibitem{Wickramasinghe}
C. C. Williams and K. Wickramasinghe, Applied Physics Letters 49, 1587 (1986)
\bibitem{Dinwiddie}
R.B. Dinwiddie, R.J. Pylkki and P.E. West, \emph{Thermal conductivity contrast imaging with a scanning thermal microscope}, Thermal conductivity 22, T.W. Wong ed, Tecnomics, Lancaster PA, 668-677, 1994
\bibitem{Millipede}
P. Vettiger, G. Cross, M. Despont, U. Drechsler, U. D\"urig, B. Gotsmann, W. Häberle, M.A. Lantz, H.E. Rothuizen, R. Stutz and G. K. Binnig, IEEE Transactions on Nanotechnology 1, 39 (2002)
\bibitem{Vettiger}
U. Drechsler, N. B\"urer, M. Despont, U. D\"urig, B. Gotsmann, F. Robin and P. Vettiger, Microelectronic Engineering 67-8, 397 (2003)
\bibitem{Yang}
J.L. Yang, M. Despont, U. Drechsler, B.W. Hoogenboom, P.L.T.M. Frederix, S. Martin, A. Engel, P. Vettiger and H.J. Hug, Applied Physics Letters 86, 134101 (2005)
\bibitem{IJHMT}
S. Lef\`evre, S. Volz and P.-O. Chapuis, International Journal of Heat and Mass Transfer 49, 251 (2006)
\bibitem{IBM}
H. F. Hamann, Y. C. Martin and H. K. Wickramasinghe, Applied Physics Letters 84, 812 (2004)
\bibitem{Majumdar}
L. Shi and A. Majumdar, Journal of Heat Transfer 124, 329 (2002)
\bibitem{Mulet2001}
J.P. Mulet, K. Joulain, R. Carminati and J.J. Greffet, Applied Physics Letters 78, 2931 (2001)
\bibitem{Bird}
G. A. Bird, \emph{Molecular Gas Dynamics and the Direct Simulation of Gas Flows}, Clarendon Press ed., Oxford, 1994
\bibitem{Rohsenow}
The accommodation factor is bigger than $\alpha=0.87$ : see p287, W. M. Rohsenow and H. Choi, \emph{Heat, Mass, and Momentun Transfer}, Prentice-Hall, 1961
\bibitem{MTE}
A similar approach for heat conduction in bulks is presented in : S. Volz, R. Carminati and K. Joulain, Microscale Thermophysical Engineering 8, 155-167 (2004)

\end{thebibliography}
\end{document}